\documentclass[apjl]{emulateapj}
\usepackage{apjfonts}
\submitted{Received 2010 June 2; accepted 2010 August 31; published --}
\journalinfo{Accepted to ApJ Letters 08/31/2010}
\usepackage{amssymb,amsmath}
\newcommand{\ha}{H$\alpha$}

\newcommand{\dg}{$^{\circ}$}
\newcommand{\Hsi}{\textit{Reuven Ramaty High Energy Solar Spectroscopic Imager}}
\newcommand{\hsi}{\textit{RHESSI}}
\newcommand{\goes}{\textit{GOES}}
\newcommand{\sm}{$\sim$}
\newcommand{\eit}{EUV Imaging Telescope}
\newcommand{\mdi}{Michelson Doppler Imager}
\newcommand{\soho}{\textit{SOHO}}

\begin{document}
\title{MOTIONS OF HARD X-RAY SOURCES DURING AN ASYMMETRIC ERUPTION}
\author{CHANG LIU\altaffilmark{1}, JEONGWOO LEE\altaffilmark{2}, JU JING\altaffilmark{1}, RUI LIU\altaffilmark{1}, NA DENG\altaffilmark{1,3}, AND HAIMIN WANG\altaffilmark{1}}
\affil{1. Space Weather Research Laboratory, New Jersey Institute of Technology, University Heights, Newark, NJ 07102-1982, USA; chang.liu@njit.edu}
\affil{2. Physics Department, New Jersey Institute of Technology, University Heights, Newark, NJ 07102-1982, USA}
\affil{3. Department of Physics and Astronomy, California State University, Northridge, CA 91330-8268, USA}
\shorttitle{MOTIONS OF HARD X-RAY SOURCES DURING AN ASYMMETRIC ERUPTION}
\shortauthors{LIU ET AL.}

\begin{abstract}
Filament eruptions and hard X-ray (HXR) source motions are commonly
observed in solar flares, which provides critical information on the
coronal magnetic reconnection. This Letter reports an event on 2005
January 15, in which we found an asymmetric filament eruption and a
subsequent coronal mass ejection together with complicated motions
of HXR sources during the GOES-class X2.6 flare. The HXR sources
initially converge to the magnetic polarity inversion line (PIL),
and then move in directions either parallel or perpendicular to the
PIL depending on the local field configuration. We distinguish the
evolution of the HXR source motion in four phases and associate each
of them with distinct regions of coronal magnetic fields as
reconstructed using a non-linear force-free field extrapolation. It
is found that the magnetic reconnection proceeds along the PIL
toward the regions where the overlying field decreases with height
more rapidly. It is also found that not only the perpendicular but
the parallel motion of the HXR sources correlates well with the HXR
lightcurve. These results are discussed in favor of the torus
instability as an important factor in the eruptive process.
\end{abstract}


\section{INTRODUCTION}
During solar flares, \ha\ ribbons form along the magnetic
polarity inversion line (PIL) and separate from each other in the direction
perpendicular to the local PIL \citep{zirin88}. This well-known
behavior has been regarded as a piece of evidence for the so-called
standard solar flare model, in which the magnetic reconnection proceeds
into the higher corona \citep[e.g.,][]{priest00}. Motions of flare ribbons parallel to the
PIL are also commonly observed but had not received much attention
until similar motions are found in hard X-ray (HXR) or EUV/UV
emissions \citep{fletcher01,grigis05b,yang09}. Since the HXR emissions
are due to high-energy particle precipitating into the chromosphere,
the parallel motion could also be associated with the primary
energy release from the corona, but its implication on solar eruptions and flare energy release is yet to be explored \citep{lee08}.

Recently, the motions of flare emissions parallel to the PIL were
related to the phenomenon called asymmetric filament
eruption \citep[e.g.,][]{tripathi06b,liur09}. During asymmetric
eruptions, only one end of the filament erupts upward with the other end
anchored, which can lead to a sequential magnetic reconnection
along the PIL and may provide an important clue to understanding the
parallel motion of HXR sources.

In this Letter, we investigate the 2005 January 15 X2.6 flare, in
which we found both the asymmetric eruption and the parallel HXR
source motion. We will discuss the implications of the HXR footpoint motions on
the eruption using a non-linear force-free field (NLFFF) extrapolation from the active region.

\section{OBSERVATIONS AND DATA REDUCTION} \label{tool}
The $\beta\delta$ active region NOAA 10720 lies close to the disk center (N15\dg, W05\dg) when the 2005 January 15 X2.6 flare peaked at 23:02~UT in \goes\ soft X-ray flux. We used high-resolution \ha~$-$~0.8~\AA\ images with a pixel scale of \sm0.6\arcsec\ and a temporal cadence of 1--2 minutes obtained with the Big Bear Solar Observatory (BBSO) to monitor the evolution of the active-region filament and that of the early flare kernels. The images of the eruptive filament as well as the surrounding coronal structure were also taken at 195~\AA\ (with 5.26\arcsec\ resolution and \sm12 minutes cadence) by the \eit\ \cite[EIT;][]{delaboudinire95}.

The evolution of the flare HXR emission was entirely registered by
\Hsi\ \citep[\hsi;][]{lin02}. CLEAN images \citep{hurford02} in the
40--100~keV energy range were reconstructed using the front segments
of detectors 3--9 (giving an FWHM resolution of \sm9.8\arcsec) with
20~s integration time, starting from
22:32:04~UT, except the time interval of $\pm$4~s at 22:34:08~UT
when the \hsi\ attenuator switched between A1 and A3 status. Considering the characteristic shape of the X-ray sources, we computed the centroid position of
each HXR source by fitting a 2D elliptical Gaussian above a minimum flux value,
and we estimated the statistical position error by varying the threshold at
40\%--60\% of the maximum flux. The footpoint velocity was then derived using
three-point Lagrangian interpolation implemented in the DERIV procedure of
IDL, as it is less sensitive to the uncertainties in the position measurement.
The total flux of each individual source was obtained by summing
up the pixel values of the entire source feature, and its
uncertainty was evaluated as $\frac{1}{3}$ of the maximum flux
outside of the source within the field of view (FOV), as implemented
in the standard \hsi\ software package.

\begin{figure}[t]
\epsscale{1.15}
\plotone{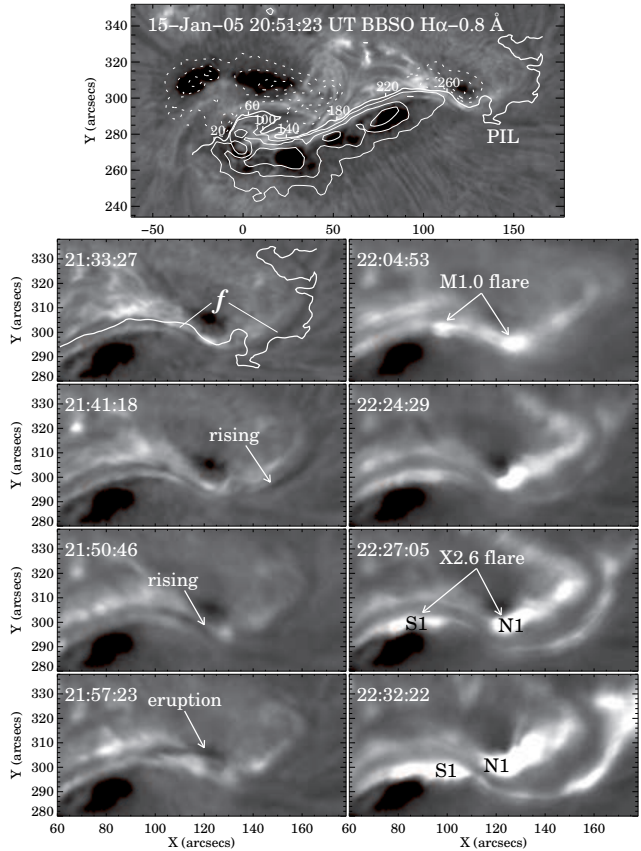}
\caption{Time sequence of BBSO \ha\ blue-wing images showing the asymmetric filament eruption. The levels of MDI contours superimposed in the top panel are $\pm$800, 1600, and 2200~G. The thick line is the main PIL. All the images presented in this paper are aligned with respect to 20:51~UT.  \label{fig_ha}}
\end{figure}

\begin{figure}[t]
\epsscale{1.15}
\plotone{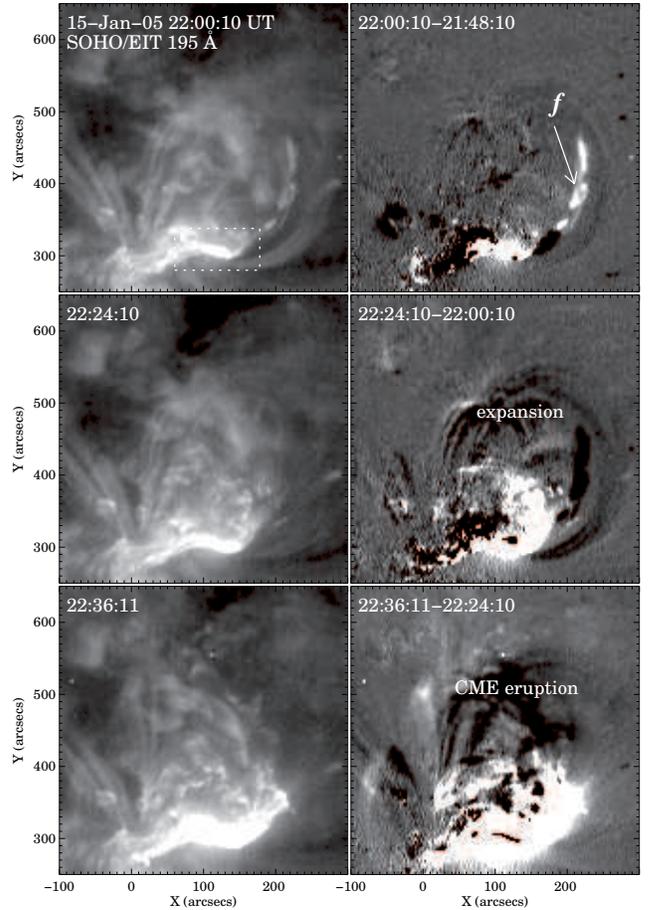}
\caption{Time sequence of \soho/EIT 195~\AA\ images showing the eruptions of the filament $f$ and the subsequent CME. The dashed box in the upper left panel indicates the FOV of the bottom 8 panels of Fig.~\ref{fig_ha}. \label{fig_eit}}
\end{figure}

\begin{figure*}[t]
\epsscale{1.1}
\plotone{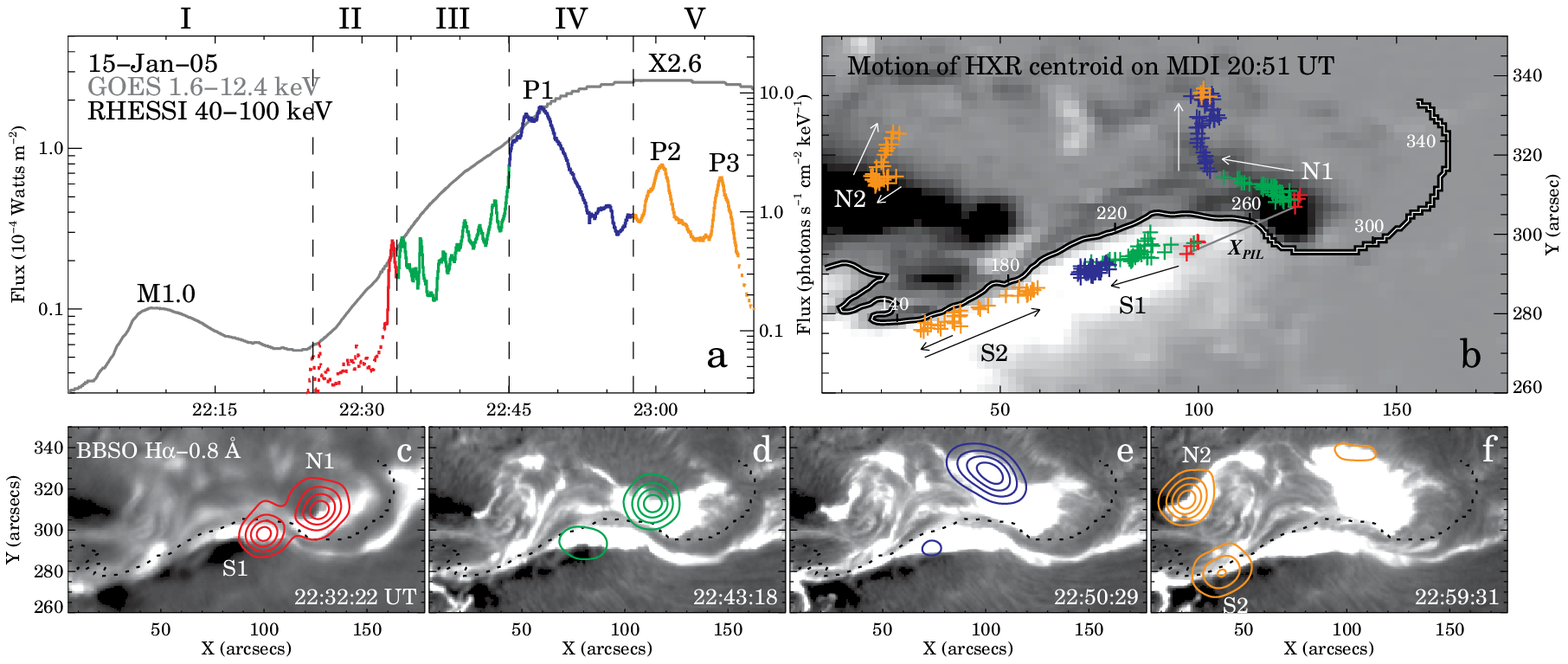}
\caption{\hsi\ HXR light curve and images together with their centroid positions ({\it pluses}) showing the event evolution during four distinct phases (II--V). Contour levels are 30\%, 50\%, 70\%, and 90\% of the maximum flux at each instance. The black bordered ($b$) and dotted ($c$--$f$) lines are the main PIL and our tracing of the filament $f$ (see Fig.~\ref{fig_ha}). See \S~\ref{HXR} for details. \label{fig_HXR}}
\end{figure*}

In this multiwavelength study, we used the full-disk 96 minute magnetogram at 20:51~UT acquired by the \mdi\ (MDI; \citealt{scherrer95}) for data alignment. The accuracy of feature matching is estimated to be about one MDI pixel size of \sm2\arcsec. Vector magnetograms of the whole active region were taken by the Digital Vector Magnetograph system at BBSO \citep{spirock02} and were calibrated using MDI level 1.8.2 data. The vector magnetogram data have been processed following \citet{jing09}: (1) the central umbral fields were filled with those measured by MDI to alleviate the influence of
polarization saturation; (2) the 180\dg\ azimuthal ambiguity in the
transverse fields was resolved using the ``minimum energy'' method
\citep{metcalf06}; (3) the projection effects were removed by
transforming the observed fields to heliographic coordinates;
and (4) the Lorentz forces and torques were minimized by a preprocessing method
\citep{wiegelmann06}. The NLFFF model was then constructed within a box
of 248~$\times$~248~$\times$~248 grid points with a size element of 1\arcsec\
using the weighted optimization method \citep{wiegelmann04} adapted to BBSO
data \citep{jing09}, the results of which show similar magnetic
structure during \sm20:30--21:30~UT. For the purpose of this study,
we analyzed the preflare NLFFF model at 20:51~UT, which was chosen for the
optimal observing condition.

\section{EVENT EVOLUTION}
Based on the HXR lightcurve, we divide the whole event into five phases, I--V (see Fig.~\ref{fig_HXR}$a$). We describe characteristic flare activities in each phase in the following subsections.

\subsection{Asymmetric Eruption (Phase I)}
Time sequence of \ha\ blue-wing images in Figure~\ref{fig_ha} shows that the western part of the filament ($f$) lying along the PIL gradually rises upward from west to east during \sm21:41--22:00~UT and subsequently erupts, while its eastern part remains undisturbed as can be seen in the time-lapse movie. This filament eruption from the western edge of the active region produced a M1.0 flare starting at 22:01~UT. The M1.0 flare appears to be a confined flare, since its flare ribbons/kernels remain almost fixed till the occurrence of the subsequent X2.6 flare at 22:24~UT (cf. images at 22:04 and 22:24~UT). The confined flaring also implies that $f$ undergoes a failed eruption \citep{schrijver09}.

In Figure~\ref{fig_eit}, we further examine the following eruption based on the evolution of overlying coronal loops shown in EUV images. Although $f$ erupts (see images at 22:00~UT), most probably it does not immediately open the overlying arcade, which expands from 22:00 to 22:24~UT (see the difference image at 22:24~UT) and finally erupts at \sm22:36~UT to become a fast halo CME \citep[also see][]{liur10}. Hence we believe that although the first eruption of $f$ (\sm21:41--22:00~UT) failed, the later eruption ($>22$:00~UT) was successful to completely tear open the overlying field, resulting in a fast CME and the X2.6 flare. This process resembles the asymmetric filament eruption reported by \citet{liur09}.

\subsection{Motion of HXR Sources of the X2.6 Flare} \label{HXR}
We show, in Figure~\ref{fig_HXR}, the evolution of the HXR footpoint sources throughout the X2.6 flare superposed on the \ha\ blue-wing images at the nearest time and the preflare MDI magnetogram. We further characterize the motion of HXR sources in Figure~\ref{fig_light} after a detailed quantitative analysis of the HXR sources.

\subsubsection{Converging Motion (Phase II)}
\begin{figure}[t]
\epsscale{1.17}
\plotone{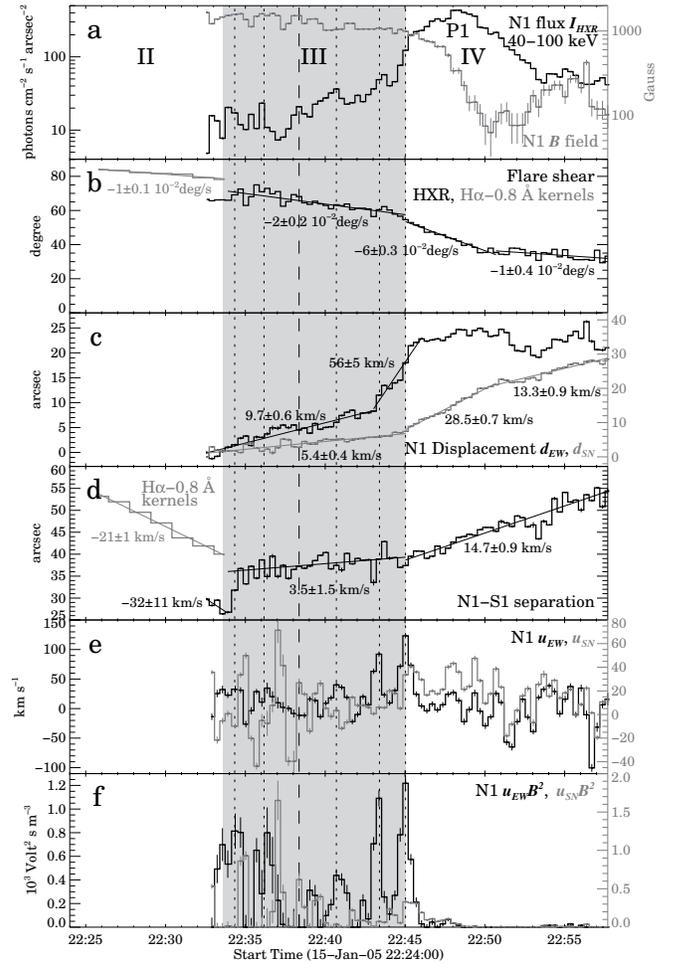}
\caption{Temporal evolution of derived parameters for the HXR sources (phase III shaded in grey). (a) HXR flux and photospheric magnetic field strength at the source centroid of N1 obtained by MDI with a largest noise level of 30~G. (b) Angle of flare shear. (c) East-west and south-north motions of N1. (d) Separation of sources N1 and S1. (e) Velocity of motion of N1. (f) Product $uB^2$. Vertical dotted and dashed lines mark the selected peaks of HXR flux. \label{fig_light}}
\end{figure}

HXR images can be reconstructed only after \sm22:32~UT. At the event onset during \sm22:25--22:33:34~UT, the flare kernels N1 and S1 show a consistent converging motion in \ha\ blue wing and HXRs (Figs.~\ref{fig_HXR}$b$$c$ and \ref{fig_light}$d$). Meanwhile, the flare shear, defined as the angle between the line connecting flare kernels and that perpendicular to the PIL \citep{ji07}, decreases steadily (Fig.~\ref{fig_light}$b$).

Similar behavior was observed before where flare kernels converge mainly in the direction perpendicular to the PIL for the entire rising phase of HXR emission \citep[e.g.,][]{ji06,ji07}. In contrast, the converging motion in the present flare is mostly parallel to the PIL, which is oriented mainly in the west-east direction (Fig.~\ref{fig_HXR}$b$).

\begin{figure*}[t]
\epsscale{1.}
\plotone{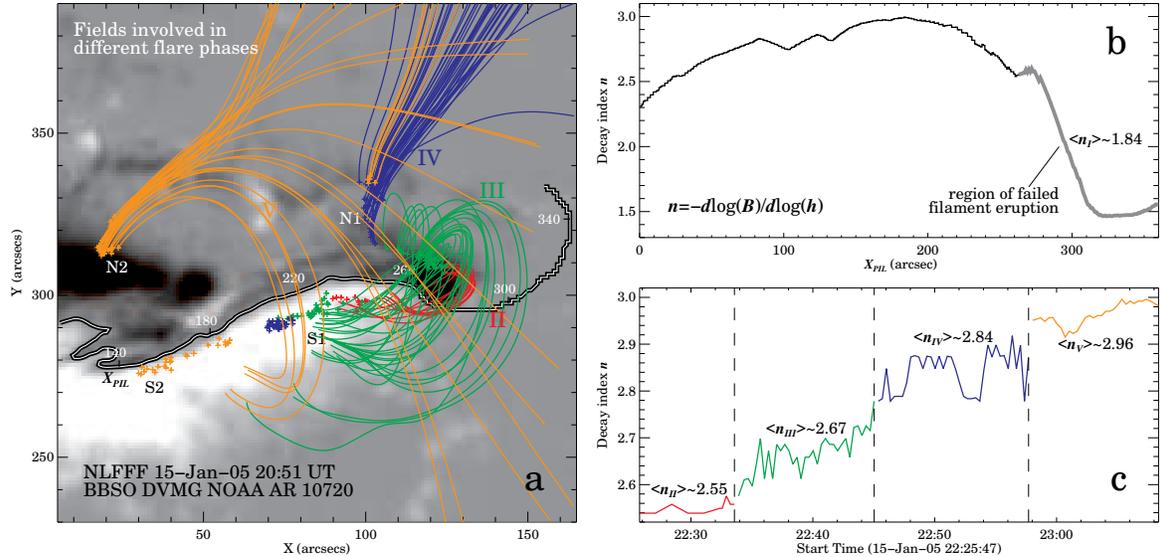}
\caption{NLFFF model computed using a preflare vector magnetogram ($a$), and the derived distribution of the decay index along the PIL/filament ($b$) and in different flare phases ($c$). See \S~\ref{NLFF} for details. \label{fig_NLFF}}
\end{figure*}

\subsubsection{Parallel Motion (Phase III)}
During 22:33:34--22:45:02~UT, N1 and S1 keep a nearly constant separation and both move preferentially parallel to the PIL (Figs.~\ref{fig_HXR}$b$$d$ and \ref{fig_light}$c$$d$). The angle of flare shear is seen to fluctuate until \sm22:38~UT but decreases overall at a higher rate than that of phase II (Fig.~\ref{fig_light}$b$). Interestingly, N1 first starts at a slower speed (\sm9.7~km~s$^{-1}$) and later the speed increases substantially (\sm56~km~s$^{-1}$) as approaching the main HXR peak P1 (Fig.~\ref{fig_light}$c$). The maximum speed reaches $\gtrsim$100~km~s$^{-1}$ (Fig.~\ref{fig_light}$e$), which is comparable to other reported values of flare kernel motions found using high-cadence \ha\ blue-wing filtergrams \citep{qiu02,lee06}.

In Figure~\ref{fig_light}$e$ and \ref{fig_light}$f$, we plot the speed $u$ and the quantity $uB^2$ of N1 in comparison with the HXR flux of N1 ($I_{HXR}$ in Fig.~\ref{fig_light}$a$).
We focus on N1, because it has the strongest HXR emission
and exhibits the most intriguing pattern of motion.
The quantity $uB^2$ is regarded as a proxy of magnetic energy release
rate under the standard 2D model (first presented by \citealt{isobe02}; see also \citealt{lee06}). We compute $B$ by averaging the four nearest pixels in the preflare MDI magnetogram around the position of N1 centroid, and we present two vector components of $u$ along the east-west and the south-north directions as $u_{EW}$ and $u_{SN}$, respectively. Remarkably, five peaks ({\it dotted lines} in Fig.~\ref{fig_light}) of $I_{HXR}$ have corresponding peaks in $u_{EW}B^2$, contributed by the footpoint motion parallel to the PIL ($u_{EW}$), while only one peak of $I_{HXR}$ at \sm22:38:22~UT ({\it dashed line} in Fig.~\ref{fig_light}) is cotemporal with that of $u_{SN}B^2$ due to the motion perpendicular to the PIL ($u_{SN}$). This peak-to-peak correspondence is significant as it remains when a three-point smoothing is applied on the displacement profile prior to the derivation of the footpoint velocity. In phase III, the calculated Pearson correlation coefficient $r$ indicates a strong correlation between $I_{HXR}$ and $u_{EW}$ with $r=0.71$ and a weak correlation between $I_{HXR}$ and $u_{EW}B^2$ with $r=0.44$, but only poor correlation is present between $I_{HXR}$ and $u_{SN}$ or $u_{SN}B^2$. Similarly for S1, the HXR flux better correlates with the parallel motion than the perpendicular motion.

As a comparison, \citet{grigis05b} also reported a motion of two HXR
footpoints  parallel to the PIL. In their event, the parallel motion
was rather smooth and no correlation between the motion and the HXR
flux was found. The authors proposed that the motion of HXR sources
simply reflects a moving trigger propagating continuously along the
PIL. \citet{lee08} presented a generalized framework in which both
the parallel and perpendicular motions represent the magnetic flux
change in the corona as a way to explain the correlation between the flux
of a coronal HXR source and parallel motions of \ha\ ribbons.
In the present case, the parallel motion of HXR footpoint
sources shows a good correlation with the HXR flux. This suggests
that the parallel motion not only maps the propagation of the
trigger but also represents the primary flare energy release.

\subsubsection{Mixed Motion (Phase IV and V)}
Phase IV includes the peak P1 of HXR emission (22:45:02--22:57:42~UT), during which N1 and S1 separate from each other (Figs.~\ref{fig_HXR}$b$$e$ and \ref{fig_light}$d$) and N1 turns northward perpendicular to the PIL (Fig.~\ref{fig_light}$c$). $I_{HXR}$ has a main contribution from the south-north motion ($u_{SN}$) and shows a moderate overall correlation with $u_{SN}B^2$ with $r=0.5$, in agreement with the standard model. The flare shear first decreases rapidly during the HXR maximum before 22:50:40~UT then slows down (Fig.~\ref{fig_light}$b$), corresponding to higher and lower northward speed of N1, respectively. However, S1 does not move systematically at this phase. Similar type of motion was found in an event studied by \citet{krucker03} in which one HXR footpoint source moves roughly along the PIL and is generally correlated with HXR flux, while other two footpoints do not show any systematic motion.

Phase V has two minor HXR peaks P2 and P3 (after 22:57:42~UT), with the conjugate footpoints, N2 and S2, showing up in distinct, eastern section of the PIL (Fig.~\ref{fig_HXR}$bf$). The motion of S2 is parallel to the PIL, while that of N2 shows both parallel and perpendicular components. A general correlation exists between all these motions with HXR flux. It is also noteworthy that the flare shear actually increases \sm30\dg\ from P2 to P3. This may be linked to harder HXR spectra of P3 than P2 found by \citet{saldanha08} for the same event.

\section{MAGNETIC FIELD STRUCTURE \label{NLFF}}
For the eruption mechanism, we consider the torus instability
(TI) the most appropriate for this event, because there was a continuous flux emergence in the region \citep{zhao08} and the supply of twisted magnetic flux into the
corona tends to cause the TI \citep[e.g.,][]{fan07}. Kink instability \citep{torok05} is excluded here because writhing motion of the filament is not apparent. Other well-known models for solar eruptions, for instance, magnetic
breakout model \citep{antiochos98} and tether-cutting reconnection
model \citep{moore01} are not considered because there are no obvious chromospheric brightenings before the eruption, either in remote regions or within the initial flaring site at the western edge of the active region.

In the TI model, the likelihood of eruption is gauged by the
decay index, defined by $n = -d {\rm log} (B) /d {\rm log} (h)$.
Here $B$ is the strength of the external field confining the erupting core field, $h$ is the height above the photosphere, and $n>1.5$--2.0 is the theoretical instability condition \citep{kliem06}. We calculate the decay index at 42--105~Mm using $B(h)$ of the NLFFF model (\S~\ref{tool}) as an approximation of the external field. Filament activation usually occurs in the similar height range according to limb observations \citep[see][and references therein]{liuy08}. We note that the NLFFF model used here is constructed at a single time in the preflare stage. Ideally we should have used time dependent NLFFF models to examine the evolving magnetic fields. To obtain reliable extrapolation, however, we cannot use flare-time magnetograms. We therefore only point out which field lines in the pre-eruption magnetic configuration are involved with the eruption in each phase. Figure~\ref{fig_NLFF}$a$ shows the magnetic field lines
stemming from N1 and N2 at their locations in each phase,
marked by different colors. We calculate $n$ at each position along the PIL with index $X_{PIL}$ (see Fig.~\ref{fig_NLFF}$a$ and also Fig.~\ref{fig_ha}
{\it top}) and plot this spatial distribution of $n$ in Figure~\ref{fig_NLFF}$b$. At each time of \ha/HXR images, we also read off, from Figure~\ref{fig_NLFF}$b$, the decay index at the particular $X_{PIL}$ where the line joining two conjugate flare footpoints intersects the PIL (see Fig.~\ref{fig_HXR}$b$). In this way we are able to approximate the decay index for the coronal region undergoing magnetic reconnection as a function of time, as shown in Figure~\ref{fig_NLFF}$c$.

In phase I, \ha\ and EUV images show that the filament $f$ at the western edge
of the active region failed to erupt, and that the associated M1.0
flare is a confined flare. At the location above $f$, the decay
index $n$ has an average value of \sm1.84. Since this value is relatively low, the failed eruption may be expected. Although it failed, the
M1.0 flare must have affected the local magnetic field in a way to
reduce the overlying magnetic flux to some extent, so that the next CME eruption can occur along with a stronger X2.6 flare. The successful CME eruption is anticipated, as the main flare phases II-V progresses from regions of $n$ about 2.5 to 3 with an average value of \sm2.7.

In phase II, low-lying and highly-sheared field lines ({\it
red} in Fig.~\ref{fig_NLFF}$a$) are involved in the reconnection process, during
which the flare kernels N1 and S1 converge with decreasing flare shear. This is consistent with the previous view that converging motion of flare kernels is due
to reconnection proceeding from more to less sheared field lines \citep{ji07}.

In phase III, N1 and S1 move parallel to the PIL from east to
west, which corresponds to the motion from strongly sheared
to less sheared fields ({\it green}) and also from stronger to weaker confinement of overlying field (Fig.~\ref{fig_NLFF}$c$). The location and direction of this HXR source motion thus support the idea that
the CME, which abruptly erupts outward at the same time \citep{liur10}, acts as the moving trigger to successively open the overlying field, in a way similar to the asymmetric filament eruption.

In Phase IV ({\it blue}), one footpoint N2 shows a standard perpendicular motion while its conjugate footpoint S2 does not. In phase V, N2 and S2 generally move in the opposite direction from west to east ({\it yellow}). Overall, phase V occurs in the region with much enhanced decay index compared to phases II--IV. In all phases, the reconnection evolves in the direction of increasing decay
index, namely, toward the region of the maximum $n\approx$ 3. It
thus appears that the magnetic reconnection tends to proceed toward
the regions with weaker magnetic confinement.

\section{SUMMARY}
We have presented a near disk-center event that shows asymmetrical filament/CME eruptions and associated flare kernel motions observed at \ha\ and HXRs, and discussed the results with the aid of NLFFF model. Major results are as follows:

1. The first activities occurred in the region of low decay index implying a strong magnetic confinement, which has produced a failed filament eruption and a relatively small flare of M1.0.

2. The subsequent main event was initiated at the same region but proceeded to nearby regions of higher decay index, i.e., weaker magnetic confinement. The eruption at this stage produced a CME and a stronger flare of X2.6 with complicated HXR footpoint motions, indicative of complex coronal magnetic structure.

3. Converging HXR footpoint motion was found in the initial stage of the X2.6 flare. As suggested before, it is considered as a result of reconnection proceeding from more to less sheared fields. In contrast to other events, the converging motion at the onset of this flare has a major component parallel to the PIL \citep[cf.][]{ji06,ji07}, and the flaring scenario does not involve quadrupolar reconnection \citep[cf.][]{ji08}.

4. During the rising phase of HXRs, footpoints moved parallel to the PIL, a phenomenon similar to what was reported by \citet{krucker03}, \citet{grigis05b}, and \citet{lee08}, and by \citet{yang09} in a statistical analysis including this event. \citet{grigis05b} suggested that the parallel HXR motion occurs because the trigger of magnetic reconnection moves along the PIL. In this case, we identified the moving trigger with the asymmetric eruption of the CME. It is further notable that the parallel HXR source motion has a peak-to-peak correlation with the HXR flux in complement to the standard 2D model.

In summary, the observations and interpretations presented in this Letter support the idea that an asymmetric eruption progressively opens the overlying field from one end of the PIL to the other end, which results in a magnetic reconnection proceeding along the PIL. A new intriguing property is that the asymmetric eruption and thus the corresponding trigger of magnetic reconnection tend to progress toward the region of weaker magnetic confinement.

\acknowledgments We thank the teams of BBSO, \hsi, and \soho\ for
excellent data set, and the referee for valuable comments that greatly improved the paper. C.L., J.J., R.L., and H.W. were supported by
NSF grants AGS 08-19662, AGS 09-36665, AGS 07-45744, and AGS
07-16950, and NASA grants NNX 08AQ90G and NNX 07AH78G. J.L. was
supported by NSF grant AST 09-08344. N.D. was supported by NASA
grant NNX 08AQ32G.

\end{document}